\documentclass[a4paper, final, 12pt]{article}
\usepackage{amsmath, amssymb, latexsym, amscd, amsthm,amsfonts,amstext}
\usepackage[mathscr]{eucal}
\usepackage{graphicx}
\usepackage{subfig}
\usepackage{float}
\usepackage{color}
\usepackage{hyperref}
\usepackage[utf8]{inputenc}
\usepackage[english]{babel}

\numberwithin{equation}{section}
\begin{document}

\title{On an $n-$Dimensional Travel Time Tomography Problem}
\author{Michael V. Klibanov}
\date{}
\maketitle

\begin{abstract}
In their seminal works Herglotz (1905) and Wiechert and Zoeppritz (1907)
have solved the so-called Travel Time Tomography Problem (TTTP) in the $1-$D
case. However, the question about stability estimates and uniqueness
theorems for an $n-$D $\left( n\geq 2\right) $ TTTP with formally determined
incomplete input data still mostly stands open after more than one hundred
years period. \textquotedblleft Formally determined input data" means that
the number $p$ of free variables in the input data equals the number $n$ of
free variables in the unknown right hand side of the governing nonliniear
eikonal PDE, $p=n$. Some previous publications demonstrate that it is
possible to develop well performed numerical methods for the TTTP with
formally determined input data, which indicates the importance of such data
for practical applications.

This is the first publication in which the above question is addressed. More
precisely, we consider a semi-discrete case, in which a PDE generated by the
eikonal equation is written in finite differences with respect to $\left(
n-1\right) $ variables. In addition, it is assumed that the solution of that
semi-discrete PDE is represented via a truncated Fourier-like series with
respect to a special orthonormal basis of functions, which depend only on
the position of the point source. Under these conditions, Lipschitz
stability estimate is proven, and this estimate implies uniqueness. An
important tool of this paper is a new Carleman estimate. Carleman Weighted
Spaces are introduced. Carleman estimates were not applied previously to
address questions about stability estimates and uniqueness theorems for the
TTTP.
\end{abstract}

\textbf{Key words}: Travel Time Tomography Problem, formally determined
data, Carleman estimate, Carleman Weighted Spaces, finite differences, a
special orthonormal basis, Lipschitz stability estimate, uniqueness.

\textbf{\ MSC code}. 35R30

\section{Introduction}

\label{sec:1}

Seminal works of Herglotz \cite{Herg} (1905) and Wiechert and Zoeppritz \cite%
{W} (1907) have provided the solution of the Travel Time Tomography Problem
(TTTP) in the $1-$D case. However, the question about stability estimates
and uniqueness theorems for the TTTP in a much more delicate $n-$D case, $%
n\geq 2,$ is still far from being fully addressed. The TTTP is the problem
of the determination of the unknown speed of propagation of seismic waves
using boundary measurements of the first times of arrivals of those waves.
This problem has broad applications in geophysics. More recently it was
noticed in \cite{KR} that the TTTP\ has applications in the inverse
scattering problem without the phase information.

In this publication we address the above mentioned open question in the $n-$%
D case with three substantially new elements listed in items 1-3 below.
Items 1-3 are not parts of the previous publications concerning stability
estimates and uniqueness theorems for the TTTP. The only exception is item 1
in the 2-D case \cite{Muh,PU}. Items 2 and 3 are not parts of \cite{Muh,PU}.
Those three items are:

\begin{enumerate}
\item The input data for our statement of the TTTP are formally determined
ones.

\item The input data for this statement are incomplete.

\item A new Carleman estimate.
\end{enumerate}

We obtain Lipschitz stability estimate and uniqueness theorem for a certain
statement of the TTTP in which items 1 and 2 are present. The TTTP is a
special Coefficient Inverse Problem (CIP), in which the unknown spatially
varying speed of propagation of seismic waves is the right hand side of the
nonlinear eikonal equation. Consider a CIP, in which the unknown coefficient
depends on $n$ free variables, and the input data depend on $p$ free
variables. The input data are called \textquotedblleft formally determined"
if $p=n$.

Our above mentioned Lipschitz stability and uniqueness results are obtained
for a semi-discrete case, i.e. under the assumption that the PDE for a
certain function generated by the solution of the governing eikonal PDE is
written in finite differences with respect to $\left( n-1\right) $
variables. In addition, we assume that this function can be represented via
a truncated Fourier-like series with respect to a special orthonormal basis.
This basis was first constructed by the author in \cite{KJIIP}. The
functions of that basis depend only on the position of the point source.

With regards to the above items 1,2 as well as to the assumptions listed in
the previous paragraph: The TTTP is long-standing and open with almost no
progress, with respect to these items, for over more than a hundred years
period. At the same time, it is shown in \cite{kin1,kin2,kin3} that the
assumptions listed lead to the developments of effective numerical methods
for the TTTP. Numerical results of \cite{kin1,kin2,kin3} demonstrate a very
good computational performances of these methods. This indicates the
importance for practical applications of Lipschitz stability and uniqueness
results for the TTTP with formally determined incomplete input data.  

With regards to the above item 3, Carleman estimates were not applied in the
past to the TTTP to address questions about stability estimates and
uniqueness theorems. In addition, to prove our new Carleman estimate, we
introduce Carleman Weighted Spaces. A similar space was introduced in \cite%
{KT} for a numerical purpose and for an elliptic partial differential
operator. The Carleman Weight Function of \cite{KT} is significantly
different from the one we use here.

Even though Lipschitz stability and uniqueness for a similar statement of
the TTTP were obtained in Theorem 7.1 of \cite{kin}, that theorem imposes
quite restrictive conditions in formulae (93) and (94) of \cite{kin}. These
formulae are used then in the second sentence of that theorem of \cite{kin}.
In contrast, the novelty of the current paper is that it is the first one
with unconditional results. These results are obtained using our novel
technique, which might also be useful in other applications. 

There are two main novel elements of our technique. First, we reduce the
target question about Lipschitz stability and uniqueness of our statement of
the TTTP\ to the same question for a boundary value problem for a system of
coupled semi-discrete PDEs of the first order with the Dirichlet boundary
condition. Second, our new Carleman estimate for the semi-discrete PDE
operator of this system ensures the Lipschitz stability and, therefore,
uniqueness of the solution of that boundary value problem. 

In their ground breaking recent work \cite{Stef} Stefanov, Uhlmann and Vasy
have considered the TTTP in the $n-$D case, with $p>n\geq 3$. Muhometov has
obtained Lipschitz stability result for the TTTP in the 2-D case with $p=n=2$
\cite{Muh}. We also refer to the work of Pestov and Uhlmann \cite{PU} with $%
p=n=2$. Muhometov and Romanov have obtained Lipschitz stability estimate for
the $n-$D case with $p>n\geq 3$ \cite{MR}, \cite[section 5 of chapter 3]{Rom}%
. The input data in all references listed in this paragraph are complete in
terms of Definition 1.1 (below).

Below $x=\left( \overline{x},z\right) \in \mathbb{R}^{n},$ $\overline{x}%
=\left( x_{1},...x_{n-1}\right) \in \mathbb{R}^{n-1},$ $z\in \mathbb{R}.$
Seismic waves are generated by point sources such as, e.g. explosions on the
surface of the Earth or earthquakes. Therefore, it is natural to assume in
the $n-$D TTTP that the input information consists of the measurements of
the first times of arrival from many sources. Let $\Omega \subset \mathbb{R}%
^{n}$ be a domain of ones interest, where that speed of the propagation of
seismic waves needs to be found. Let $\partial \Omega $ be the piecewise
smooth boundary of $\Omega ,$ let $S\subseteq \partial \Omega $ be a part of
that boundary. Let $L\subset \mathbb{R}^{n},$ $L\in C^{1}$ be an $s-$%
dimensional piecewise smooth manifold, where $s\in \left[ 1,n-1\right] $. We
assume that measurements of the first times of arrival of waves are carried
out for all $x\in S$ and for all locations of the point source $x_{0}\in L.$
Obviously $S$ is an $\left( n-1\right) -$D manifold. Hence, the measured
data depend on $n-1+s=p$ free variables. Therefore, the data are formally
determined ones only if $s=1$, i.e. if $L$ is a curve. Thus, the input data
for the 2-D TTTP\ are always formally determined ones \cite{Muh,PU}.
However, if $n\geq 3$ and $s\geq 2,$ then the number of free variables in
the data is $n-1+s=p>n,$ which means that such data are not formally
determined ones. We assume in all our derivations below that $L$ is a
segment of a straight line, which means that $s=1,$ and, therefore, $%
n-1+s=p=n$.

\textbf{Definition 1.1.} Let $L\subset \mathbb{R}^{n}$ \emph{be the above
mentioned manifold}. \emph{We call the input data for the TTTP incomplete if 
}$S\neq \partial \Omega $\emph{\ and }$L\neq \partial \Omega $\emph{. If,
however, }$S=L=\partial \Omega $ \emph{then we call such input data
complete. }

We develop here a version of the method, which was first introduced in the
field of CIPs in \cite{BukhKlib}. This method is based on Carleman
estimates. The idea of \cite{BukhKlib} has generated many publications of
many authors. Since this paper is not a survey, we refer only to some of
those \cite{Bell,Hamr,Isakov,Ksurvey,KL,KL1,Liu} and references cited
therein. We also point out that Carleman estimates are quite effectively
used for proofs of unique continuation results. In this regard we refer to
two distinguished works of Ionescu and Klainerman \cite{IK1,IK2}.

This paper is organized as follows. In section 2 we present our statement of
the Travel Time Tomography Problem. In section 3 we describe the above
mentioned orthonormal basis. In section 4 we describe our transformation
procedure. In particular, we introduce in this section partial finite
differences with respect to $\left( n-1\right) $ variables. This
transformation procedure of section 4 allows us to formulate and prove our
Lipschitz stability estimate and uniqueness theorem, which are our target
results. This is done in Theorem 5.1 of section 5. In addition, we formulate
and prove in section 5 Theorem 5.2, which is a new Carleman estimate. All
functions considered below are real valued ones.

\section{Statement of the Travel Time Tomography Problem (TTTP)}

\label{sec:2}

We present in this section a statement of the TTTP with formally determined
incomplete data. Denote $n\left( x\right) $ the refractive index of the
medium. Then $c\left( x\right) =1/n\left( x\right) $ is the speed of the
propagation of waves. Let $m\left( x\right) =n^{2}\left( x\right) .$

\subsection{The Domain}

\label{sec:2.1}

Although the technique of this paper works for rather general domains $%
\Omega \subset \mathbb{R}^{n},$ to simplify the presentation, we consider
only one domain: a rectangular prism in $\mathbb{R}^{n}$.

First, let $A,B>0$ be two numbers and let the number $d\in \left(
-A,A\right) .$ We define the domain $\Omega \subset \mathbb{R}^{n}$ as:%
\begin{equation}
\Omega =\left\{ x=\left( \overline{x},z\right) :-A<x_{i}<A,z\in \left(
0,B\right) ,i=1,...,n-1\right\} .  \label{2.4}
\end{equation}%
We now introduce a segment of a straight line along which the point source
runs as:%
\begin{equation}
\left. L=\left\{ 
\begin{array}{c}
\left\{ 
\begin{array}{c}
x_{0}=\left( \overline{x},b\right) :x_{1}=\alpha \in \left( -A,A\right) , \\ 
x_{2},...,x_{n-1}=d\in \left( -A,A\right) ,\text{ }z=b<0,%
\end{array}%
\right\} \text{ if }n\geq 3, \\ 
\\ 
\left\{ x_{0}=\left( x_{1},b\right) :x_{1}=\alpha \in \left( -A,A\right)
,z=b<0\right\} ,\text{ if }n=2,%
\end{array}%
\right. \right.  \label{2.6}
\end{equation}%
where $\alpha \in \left( -A,A\right) $ is a running parameter and $b<0$ is a
fixed number. Obviously $L\cap \overline{\Omega }=\varnothing .$

Let $\overline{m}>0$ be a given number. We assume that%
\begin{equation}
m\left( x\right) \geq \overline{m},\quad x\in \mathbb{R}^{n},  \label{2.2}
\end{equation}%
\begin{equation}
m\in C^{1}\left( \mathbb{R}^{n}\right) .  \label{2.3}
\end{equation}%
Based on (\ref{2.4}) and (\ref{2.6}), we impose, in addition to (\ref{2.2}),
(\ref{2.3}), the following conditions on the function $m\left( x\right) $ : 
\begin{equation}
m\left( x\right) =1\text{ for }x\in \left\{ z<0\right\} \cup \left\{
\left\vert x_{i}\right\vert >A,\text{ }i=1,...,n-1\right\} ,  \label{2.8}
\end{equation}%
\begin{equation}
\partial _{z}m\left( x\right) \geq 0,\text{ }x\in \mathbb{R}^{n}.
\label{2.9}
\end{equation}

The monotonicity condition (\ref{2.9}) is not a too restrictive one. Indeed,
see, e.g. formulas (3.24) and (3.24$^{\prime }$) in chapter 3 of the book 
\cite{Rom}. In addition, a monotonicity condition is a part of \ works of
Herglotz and an analogous monotonicity condition was actually imposed in the
1D case in the originating classical works of Herglotz and Wiechert and
Zoeppritz \cite{Herg,W}. Their method is described in \cite[section 3 of
chapter 3]{Rom}. The same is true for two other monotonicity conditions
described below.

The function $m\left( x\right) $ generates the Riemannian metric 
\begin{equation*}
d\tau =\sqrt{m(x)}\left\vert dx\right\vert .
\end{equation*}

\textbf{Regularity Assumption.} \emph{We assume the regularity of geodesic
lines. In other words, we assume that each pair of points }$x\in \overline{%
\Omega }$\emph{\ and }$x_{0}\in L$\emph{\ can be connected by a single
geodesic line }$\Gamma \left( x,x_{0}\right) .$

Everywhere below we rely on this assumption without mentioning it. This
assumption is used quite often in the studies of the TTTP, see, e.g. \cite%
{Muh,MR,PU,Rom}. A sufficient condition guaranteeing the regularity was
derived in \cite{Rom2}. Temporary denote $x=\left( x_{1},...,x_{n}\right)
\in \mathbb{R}^{n}.$ Assuming that $m\left( x\right) \in C^{2}\left( \mathbb{%
R}^{n}\right) ,$ this condition is%
\begin{equation*}
\sum\limits_{i.j=1}^{n}\frac{\partial ^{2}\ln \left( m\left( x\right)
\right) }{\partial x_{j}\partial x_{i}}\xi _{i}\xi _{j}\geq 0,\text{ }%
\forall \xi \in \mathbb{R}^{n},\text{ }\forall x\in \mathbb{R}^{n}.
\end{equation*}

Let $\tau \left( x,x_{0}\right) $ be the travel time, which the wave,
generated at the source $x_{0}\in L,$ needs to travel to the point $x\in 
\overline{\Omega }.$ Then \cite[chapter 3]{Rom}:%
\begin{equation*}
\tau \left( x,x_{0,1}\right) =\int\limits_{\Gamma \left( x,x_{0}\right) }%
\sqrt{m\left( x\left( s\right) \right) }{ds},\text{ }\forall \left(
x,x_{0}\right) \in \overline{\Omega }\times L,
\end{equation*}%
where $\Gamma \left( x,x_{0}\right) $ is the unique geodesic line connecting
points $x_{0,1}$ and $x$ and $ds$ is the euclidean arc length. The function $%
\tau \left( x,x_{0}\right) $ satisfies the eikonal equation, 
\begin{equation}
\left( \nabla _{x}\tau \right) ^{2}=m\left( x\right) ,\text{ }x\mathbf{\in }%
\mathbb{R}^{n},  \label{2.1}
\end{equation}%
\begin{equation}
\tau \left( x,x_{0}\right) =O\left( \left\vert x-x_{0}\right\vert \right) ,%
\text{ }{\text{a}}\text{{s}}{\ }\left\vert x-x_{0}\right\vert \rightarrow 0.
\label{2.01}
\end{equation}%
Note that (\ref{2.4}), (\ref{2.6}), (\ref{2.8}), (\ref{2.1}) and (\ref{2.01}%
) imply 
\begin{equation}
\left. \tau \left( x,x_{0}\right) =\left\{ 
\begin{array}{c}
\left[ \left( x_{1}-\alpha \right) ^{2}+\sum\limits_{i=2}^{n-1}\left(
x_{i}-d\right) ^{2}+\left( z-b\right) ^{2}\right] ^{1/2},\alpha \in \left(
-A,A\right) , \\ 
\text{for }z\in \left( b,0\right) \text{ if }n\geq 3, \\ 
\left[ \left( x_{1}-\alpha \right) ^{2}+\left( z-b\right) ^{2}\right] ^{1/2},%
\text{ }\alpha \in \left( -A,A\right) ,\text{ } \\ 
\text{for }z\in \left( b,0\right) \text{ if }n=2.%
\end{array}%
\right. \right.  \label{2.14}
\end{equation}%
Using (\ref{2.6}), denote%
\begin{equation}
\tau \left( x,x_{0}\right) =\tau \left( x,\alpha \right) ,\text{ }x\in 
\mathbb{R}^{n},\text{ }\alpha \in \left( -A,A\right) .  \label{2.15}
\end{equation}%
We assume below that there exist the following derivatives of the function $%
\tau \left( x,\alpha \right) :$ 
\begin{equation}
\left. 
\begin{array}{c}
\partial _{x_{i}}^{r}\partial _{\alpha }^{j}\tau \left( x,\alpha \right)
,\partial _{z}^{r}\partial _{\alpha }^{j}\tau \left( x,\alpha \right)
,\partial _{x_{i}}\partial _{z}\partial _{\alpha }^{j}\tau \left( x,\alpha
\right) \in C\left( \overline{\Omega }\times \left[ -A,A\right] \right) , \\ 
\text{ } \\ 
i=1,...,n-1,\text{ }r=0,1,2,\text{ }j=0,1.%
\end{array}%
\right.  \label{2.150}
\end{equation}%
As to (\ref{2.150}), we note that some extra smoothness requirements are
traditionally of a secondary concern in the theory of CIPs, see, e.g. \cite%
{Nov,Rom1}.

\textbf{Lemma 2.1. }\emph{Assume that conditions (\ref{2.4})-(\ref{2.9}) and
(\ref{2.15}) are in place. Then there exists a number }$c=c\left( n,%
\overline{m},b,\Omega \right) >0$\emph{\ depending only on listed parameters
such that}%
\begin{equation}
\partial _{z}\tau \left( x,\alpha \right) \geq c,\text{ }\forall x\mathbf{=}%
\left( \overline{x},z\right) \in \Omega ,\text{ }\forall x_{0}\in L.
\label{2.16}
\end{equation}

Lemma 2.1 is the $n-$D analog of Lemma 5.1 of \cite{KR1}, which was proven
in the 3-D case, also, see Lemma 3.1 in \cite{kin1}. We omit the proof for
the $n-$D case since this proof is almost identical with one in \cite{KR1}.

\subsection{Statement of the TTTP}

\label{sec:2.2}

By (\ref{2.4}) the boundary $\partial \Omega $ of the domain $\Omega $
consists of three parts:%
\begin{equation}
\left. 
\begin{array}{c}
\partial \Omega _{1}=\partial _{1}\Omega \cup \partial _{2}\Omega \cup
\partial _{3}\Omega , \\ 
\\ 
\partial _{1}\Omega =\left\{ 
\begin{array}{c}
x=\left( x_{1},x_{2},...,x_{n-1},z\right) : \\ 
-A<x_{i}<A,\text{ }i=1,...,n-1,\text{ }z=B%
\end{array}%
\right\} , \\ 
\\ 
\partial _{2}\Omega =\left\{ 
\begin{array}{c}
x=\left( x_{1},x_{2},...,x_{n-1},z\right) : \\ 
x=\left( x_{1},x_{2},...,x_{n-1},z\right) : \\ 
-A<x_{i}<A,\text{ }i=1,...,n-1,\text{ }z=0%
\end{array}%
\right\} , \\ 
\\ 
\partial _{3}\Omega =\partial \Omega \diagdown \left( \partial _{1}\Omega
\cup \partial _{2}\Omega \right) .%
\end{array}%
\right.  \label{2.28}
\end{equation}%
It follows from (\ref{2.14}), (\ref{2.15}) and (\ref{2.28}) that the
function $\tau \left( x\mathbf{,}\alpha \right) $ is known for $x\in
\partial _{1}\Omega ,\alpha \in \left( -A,A\right) .$ Therefore, we assume
in our TTTP that this function is unknown for $x\in \partial _{3}\Omega $
and known for $x\in \partial _{1}\Omega .$ We use notation (\ref{2.15}).

\textbf{Travel Time Tomography Problem (TTTP).} \emph{Assume that
conditions\ (\ref{2.4})-(\ref{2.9}) and (\ref{2.28}) hold. Suppose that the
function }$m\left( x\right) $\emph{\ is unknown for }$x\in \Omega _{1}.$%
\emph{\ Find }$m\left( x\right) $\emph{\ for }$x\in \Omega ,$\emph{\
assuming that the following two functions are known }$f_{1}\left( \overline{x%
},\alpha \right) $ and $f_{2}\left( x,\alpha \right) ,$ where 
\begin{equation}
\tau \left( \overline{x}\mathbf{,}B,\alpha \right) =f_{1}\left( \overline{x}%
,\alpha \right) ,\text{ }\forall \left( \overline{x}\mathbf{,}B\right) \in
\partial _{1}\Omega ,\text{ }\forall \alpha \in \left( -A,A\right) .
\label{2.29}
\end{equation}%
\begin{equation}
\tau \left( x,\alpha \right) =f_{2}\left( x,\alpha \right) ,\text{ }\forall
\left( x,\alpha \right) \in \partial _{3}\Omega \times \left( -A,A\right) .
\label{2.30}
\end{equation}

In terms of Definition 1.1, the data for our TTTP are incomplete since $L$
is an interval of a straight line.

\section{Orthonormal Basis}

\label{sec:3}

The orthonormal basis described in this section was first introduced in \cite%
{KJIIP}, also, see the book \cite[section 6.2.3]{KL}. Previously this basis
was applied only for some numerical methods in, e.g. \cite{Klikin,KL} as
well as in a number of other publications. The current paper is the first
one, where this basis is used for proofs of stability and uniqueness
theorems for a CIP.

Consider the set of functions $\{\alpha ^{n}e^{\alpha }\}_{n=0}^{\infty
}\subset L_{2}(-A,A).$ This is a set of linearly independent functions,
which is complete in the space $L_{2}(-A,A)$. Applying the Gram-Schmidt
orthonormalization procedure to this set we obtain the orthonormal basis $%
\{\Psi _{j}\left( \alpha \right) \}_{j=0}^{\infty }$ in $L_{2}(-A,A)$. For
any $j\geq 0$ the function $\Psi _{j}(\alpha )$ of this basis, has the form 
\begin{equation}
\Psi _{j}(\alpha )=P_{j}(\alpha )e^{\alpha },  \label{3.1}
\end{equation}%
where $P_{j}(\alpha )$ is a polynomial of the degree $j$. Theorem 3.1 is the
key result for this basis.

\textbf{Theorem 3.1} (\cite{KJIIP}, \cite[Theorem 6.2.1]{KL}). \emph{Denote }%
$\left( ,\right) $\emph{\ the scalar product in }$L_{2}(-A,A)$\emph{. Let }%
\begin{equation}
a_{j_{1},j_{2}}=\left( \Psi _{j_{1}}^{\prime },\Psi _{j_{2}}\right)
=\int\limits_{-A}^{A}\Psi _{j_{1}}^{\prime }\left( \alpha \right) \Psi
_{j_{2}}\left( \alpha \right) d\alpha .  \label{3.2}
\end{equation}%
\emph{\ Then}%
\begin{equation}
a_{j_{1},j_{2}}=\left\{ 
\begin{array}{c}
1\text{ if }j_{1}=j_{2}, \\ 
0\text{ if }j_{2}>j_{1}.%
\end{array}%
\right.  \label{3.3}
\end{equation}%
\emph{Let }$N\geq 1$\emph{\ be an integer. Consider the }$N\times N$\emph{\
matrix }%
\begin{equation}
M_{N}=\left( a_{_{j_{1},j_{2}}}\right) _{\left( j_{1},j_{2}\right) =\left(
0,0\right) }^{\left( N-1,N-1\right) }.  \label{3.4}
\end{equation}%
\emph{\ Then by (\ref{3.3}) }$\det M_{N}=1.$\emph{\ Thus, the matrix }$M_{N}$%
\emph{\ is invertible with its inverse }$M_{N}^{-1}$\emph{.}

We observe that analogs of the matrix $M_{N}$ for classical orthonormal
polynomials and for the basis of trigonometric functions do not have
inverses since the first function in both cases is a constant with its
derivative being identical zero.

\section{Transformations}

\label{sec:4}

\subsection{The difference of two solutions}

\label{sec:4.1}

Consider two functions $m_{1},m_{2}$ satisfying conditions (\ref{2.2})-(\ref%
{2.9}). Let $\tau _{1}\left( x,\alpha \right) $ and $\tau _{2}\left(
x,\alpha \right) $ be two corresponding solutions of eikonal equation (\ref%
{2.1}) with condition (\ref{2.01}). Let $f_{1,1}\left( \overline{x},\alpha
\right) $ and $f_{1,2}\left( \overline{x},\alpha \right) $ be two
corresponding functions (\ref{2.29}). Denote%
\begin{equation}
\left. 
\begin{array}{c}
\widetilde{m}\left( x\right) =m_{1}\left( x\right) -m_{2}\left( x\right) ,%
\text{ }\widetilde{\tau }\left( x,\alpha \right) =\text{ }\tau _{1}\left(
x,\alpha \right) -\text{ }\tau _{2}\left( x,\alpha \right) , \\ 
\\ 
p\left( x,\alpha \right) =\tau _{1}\left( x,\alpha \right) +\text{ }\tau
_{2}\left( x,\alpha \right) , \\ 
\\ 
\widetilde{f}\left( \overline{x},\alpha \right) =f_{1,1}\left( \overline{x}%
,\alpha \right) -f_{1,2}\left( \overline{x},\alpha \right) .%
\end{array}%
\right.  \label{4.1}
\end{equation}%
In addition, let $f_{2,1}\left( x,\alpha \right) $ and $f_{2,2}\left(
x,\alpha \right) $ be two corresponding boundary functions (\ref{2.30}). We
assume that%
\begin{equation}
f_{2,1}\left( x,\alpha \right) -f_{2,2}\left( x,\alpha \right) =0,\text{ }%
\forall \left( x,\alpha \right) \in \partial _{3}\Omega \times \left(
-A,A\right) .  \label{4.01}
\end{equation}

Using (\ref{2.1}), (\ref{2.14}), (\ref{2.15}), (\ref{2.150})-(\ref{2.30}), (%
\ref{4.1}) and (\ref{4.01}), we obtain%
\begin{equation}
\text{ }\widetilde{\tau }_{z}p_{z}+\sum\limits_{i=1}^{n-1}\widetilde{\tau }%
_{x_{i}}p_{x_{i}}=\widetilde{m}\left( x\right) ,\text{ }x\in \Omega ,\text{ }%
\alpha \in \left( -A,A\right) ,  \label{4.2}
\end{equation}%
\begin{equation}
\text{ }\widetilde{\tau }\left( \overline{x},B,\alpha \right) =\widetilde{f}%
\left( \overline{x},\alpha \right) ,\text{ }\left( \overline{x}\mathbf{,}%
B\right) \in \partial _{1}\Omega ,\text{ }\alpha \in \left( -A,A\right) ,
\label{4.3}
\end{equation}%
\begin{equation}
\widetilde{\tau }\left( \overline{x},0,\alpha \right) =0,\text{ }\left( 
\overline{x}\mathbf{,}0\right) \in \partial _{2}\Omega ,\text{ }\alpha \in
\left( -A,A\right) ,  \label{4.4}
\end{equation}%
\begin{equation}
\widetilde{\tau }\left( x,\alpha \right) =0,\text{ }\forall \left( x,\alpha
\right) \in \partial _{3}\Omega \times \left( -A,A\right) .  \label{4.40}
\end{equation}%
Similarly with \cite[formula (3.1)]{kin1}, denote 
\begin{equation}
u\left( x,\alpha \right) =\text{ }\widetilde{\tau }\left( x,\alpha \right)
p_{z}\left( x,\alpha \right) ,\text{ }x\in \Omega ,\text{ }\alpha \in \left(
-A,A\right) .  \label{4.5}
\end{equation}%
Using Lemma 2.1 and the second line of (\ref{4.1}), we obtain 
\begin{equation}
p_{z}\left( x,\alpha \right) \geq 2c.  \label{4.6}
\end{equation}%
Hence, (\ref{4.5}) implies 
\begin{equation}
\widetilde{\tau }\left( x,\alpha \right) =\frac{u\left( x,\alpha \right) }{%
p_{z}\left( x,\alpha \right) }.  \label{4.7}
\end{equation}%
Hence, equation (\ref{4.2}) becomes 
\begin{equation}
u_{z}+\sum\limits_{i=1}^{n-1}q_{i}\left( x,\alpha \right)
u_{x_{i}}+q_{0}\left( x,\alpha \right) u=\widetilde{m}\left( x\right) ,\text{
}x\in \Omega ,\alpha \in \left( -A,A\right) ,  \label{4.8}
\end{equation}%
where by (\ref{2.150}) and (\ref{4.7}) 
\begin{equation}
q_{i}\left( x,\alpha \right) =\frac{p_{x_{i}}}{p_{z}}\left( x,\alpha \right)
,\text{ }q_{0}\left( x,\alpha \right) =-\sum\limits_{i=1}^{n-1}\frac{%
p_{x_{i}z}p_{x_{i}}}{p_{z}^{2}}\left( x,\alpha \right) -\frac{p_{zz}}{%
p_{z}^{2}}\left( x,\alpha \right) .  \label{4.9}
\end{equation}%
In addition, by (\ref{4.3})-(\ref{4.5})%
\begin{equation}
\left. 
\begin{array}{c}
u\left( \overline{x},B,\alpha \right) =g\left( \overline{x},\alpha \right) ,%
\text{ }\left( \overline{x}\mathbf{,}B\right) \in \partial _{1}\Omega ,\text{
}\alpha \in \left( -A,A\right) , \\ 
\\ 
u\left( \overline{x},0,\alpha \right) =0,\text{ }\left( \overline{x}\mathbf{,%
}0\right) \in \partial _{2}\Omega ,\text{ }\alpha \in \left( -A,A\right) ,
\\ 
\\ 
u\left( x,\alpha \right) =0,\text{ }\forall \left( x,\alpha \right) \in
\partial _{3}\Omega \times \left( -A,A\right) .%
\end{array}%
\right. \text{ }  \label{4.10}
\end{equation}%
where%
\begin{equation}
g\left( \overline{x},\alpha \right) =\widetilde{f}\left( \overline{x},\alpha
\right) p_{z}\left( \overline{x},B,\alpha \right) ,\text{ }\left( \overline{x%
}\mathbf{,}B\right) \in \partial _{1}\Omega ,\text{ }\alpha \in \left(
-A,A\right) .  \label{4.010}
\end{equation}%
In addition, by (\ref{2.150}), (\ref{4.1}), (\ref{4.6}) and (\ref{4.9})-(\ref%
{4.010}) 
\begin{equation}
\left. 
\begin{array}{c}
\partial _{\alpha }^{j}g\in C\left( \partial _{1}\Omega \times \left[ -A,A%
\right] \right) ,\text{ }j=0,1, \\ 
\\ 
\partial _{\alpha }^{j}q_{i}\in C\left( \overline{\Omega }\times \left[ -A,A%
\right] \right) ,\text{ }j=0,1;\text{ }i=0,...,n-1.%
\end{array}%
\right.  \label{4.0100}
\end{equation}%
Recall that in (\ref{2.16}) the number $c=c\left( n,\overline{m},b,\Omega
\right) >0$. Hence, (\ref{2.150}), (\ref{4.3}), (\ref{4.5})-(\ref{4.7}) and (%
\ref{4.9})-(\ref{4.0100}) imply that there exists a number $D$ such that 
\begin{equation}
\left. 
\begin{array}{c}
D=D\left( n,\overline{m},b,\Omega \right) >0, \\ 
\\ 
\left\Vert \partial _{\alpha }^{j}g\right\Vert _{C\left( \partial _{1}\Omega
\times \left[ -A,A\right] \right) }\leq D,\text{ }\left\Vert \partial
_{\alpha }^{j}q_{i}\right\Vert _{C\left( \overline{\Omega }\times \left[ -A,A%
\right] \right) }\leq D,\text{ } \\ 
\\ 
j=0,1;\text{ }i=0,...n-1.%
\end{array}%
\right.  \label{4.11}
\end{equation}

Following the first step of the method of \cite{BukhKlib}, we now
differentiate both sides of equation (\ref{4.8}) with respect to $\alpha ,$
also, see, e.g. \cite[formula (3.6)]{kin1} for a similar step. Since $%
\partial _{\alpha }\widetilde{m}\left( x\right) \equiv 0,$ then, using (\ref%
{4.11}), we obtain%
\begin{equation}
\left. 
\begin{array}{c}
\partial _{\alpha }u_{z}+\sum\limits_{i=1}^{n-1}q_{i}\left( x,\alpha
\right) \partial _{\alpha }u_{x_{i}}+q_{0}\left( x,\alpha \right) \partial
_{\alpha }u+ \\ 
\\ 
+\sum\limits_{i=1}^{n-1}\partial _{\alpha }q_{i}\left( x,\alpha \right)
u_{x_{i}}+\partial _{\alpha }q_{0}\left( x,\alpha \right) u=0,\text{ }x\in
\Omega ,\alpha \in \left( -A,A\right) .%
\end{array}%
\right.  \label{4.100}
\end{equation}

Equation (\ref{4.100}) contains both the function $u\left( x,\alpha \right) $
and its $\alpha -$derivative $u_{\alpha }\left( x,\alpha \right) .$ In the
case of CIPs for time dependent PDEs, to use the method of \cite{BukhKlib},
one would express the original function via its time derivative using the
Fundamental Theorem of Calculus and the initial condition for the original
function, see, e.g. \cite[lines between (3.15) and (3.16)]{Ksurvey}. In our
case, the latter would mean%
\begin{equation}
u\left( x,\alpha \right) =\int\limits_{\alpha _{0}}^{\alpha }u_{\alpha
}\left( x,\beta \right) d\beta +u\left( x,\alpha _{0}\right) ,\text{ }\left(
x,\alpha \right) \in \Omega \times \left( -A,A\right) ,  \label{4.12}
\end{equation}%
where $\alpha _{0}\in \left[ -A,A\right] $ is a fixed number. However, the
function $u\left( x,\alpha _{0}\right) $ is unknown for any value of $\alpha
_{0}\in \left[ -A,A\right] $. Hence, formula (\ref{4.12}) is inapplicable in
our case. This is the reason why do we use below orthonormal basis (\ref{3.1}%
).

\subsection{Partial finite differences}

\label{sec:4.2}

In this subsection, we rewrite equation (\ref{4.100}) via partial finite
differences with respect to $x_{1},x_{2},...,x_{n-1}$. Let $k\geq 2$ be an
integer. Keeping in mind (\ref{2.4}), consider $n-1$ partitions of the
interval $\left( -A,A\right) $ with the grid step size $h$:%
\begin{equation}
\left. 
\begin{array}{c}
A=x_{i,0}<x_{i,1}<...<x_{i,k}=A,\text{ }x_{i,j+1}-x_{i,j}=h, \\ 
\text{ } \\ 
i=1,...,n-1\text{, }j=0,...,k-1.%
\end{array}%
\right.  \label{4.13}
\end{equation}%
Let $h_{0}\in \left( 0,1\right) $ be a fixed number. We assume below that 
\begin{equation}
h\in \left[ h_{0},1\right) .  \label{4.14}
\end{equation}%
Define the discrete subset $\Omega ^{h}\hspace{0.5em}$of the domain $\Omega $
as:%
\begin{equation}
\left. \Omega ^{h}=\left\{ 
\begin{array}{c}
x^{h}=\left( \overline{x}^{h},z\right) =\left\{ x_{1,j_{1}},...,x_{\left(
n-1\right) ,j_{n-1}},z\right\} , \\ 
\\ 
j_{1},...,j_{n-1}\in \left[ 1,k-1\right] ,z\in \left( 0,B\right)%
\end{array}%
\right\} \subset \Omega .\right.  \label{4.15}
\end{equation}%
Let $\partial \Omega ^{h}\subset \partial \Omega $ be the boundary of $%
\Omega ^{h}.$ Using (\ref{2.28}) and (\ref{4.13})-(\ref{4.15}), denote the
semi-discrete analogs of parts $\partial _{1}\Omega ,\partial _{2}\Omega $
and $\partial _{3}\Omega $ of $\partial \Omega ^{h}$ as: 
\begin{equation}
\left. 
\begin{array}{c}
\partial _{1}\Omega ^{h}=\left\{ 
\begin{array}{c}
x^{h}=\left( \overline{x}^{h},B\right) =\left\{ x_{1,j_{1}},...,x_{\left(
n-1\right) ,j_{n-1}},B\right\} : \\ 
\\ 
-A<x_{ij_{i}}<A,i=1,...,n-1,\text{ }j_{i}=1,...k-1%
\end{array}%
\right\} \subset \partial _{1}\Omega , \\ 
\\ 
\partial _{2}\Omega ^{h}=\left\{ 
\begin{array}{c}
x^{h}=\left( \overline{x}^{h},0\right) =\left\{ x_{1,j_{1}},...,x_{\left(
n-1\right) ,j_{n-1}},0\right\} : \\ 
\\ 
-A<x_{ij_{i}}<A,\text{ }i=1,...,n-1,\text{ }j_{i}=1,...k-1%
\end{array}%
\right\} \subset \partial _{2}\Omega , \\ 
\\ 
\partial _{3}\Omega ^{h}=\partial \Omega ^{h}\diagdown \left( \partial
_{1}\Omega ^{h}\cup \partial _{2}\Omega ^{h}\right) .%
\end{array}%
\right.  \label{4.16}
\end{equation}

Let the function $P(x,\alpha )\in C^{1}(\overline{\Omega }\times \left[ -A,A%
\right] )$. Then $P^{h}(x^{h},\alpha )=P^{h}(\overline{x}^{h},z,\alpha )$ is
its semi-discrete analog defined on the set $\overline{\Omega }^{h}\times %
\left[ -A,A\right] $,%
\begin{equation}
P^{h}(\overline{x}^{h},z,\alpha )=P(\overline{x}^{h},z,\alpha ),\text{ }%
\forall (\overline{x}^{h},z,\alpha )\in \overline{\Omega }^{h}\times \left[
-A,A\right] .  \label{4.17}
\end{equation}%
We now define the finite difference derivatives of the semi-discrete
function $P^{h}(\overline{x}^{h},z,\alpha )$ with respect to variables $%
x_{i},i=1,...,n-1$ at interior points of $\Omega ^{h}$. Consider an
arbitrary point $\left( \overline{x}^{h},z,\alpha \right) \in \Omega
^{h}\times \left( -A,A\right) .$ It follows from (\ref{4.15}) that there
exist integers $j_{1},...,j_{i},...,j_{n-1}\in \left[ 1,k-1\right] $ such
that%
\begin{equation}
\left( \overline{x}^{h},z,\alpha \right) =\left(
x_{1,j_{1}},...,x_{i,j_{i}},...,x_{\left( n-1\right) ,j_{n-1},}z,\alpha
\right) \in \Omega ^{h}\times \left( -A,A\right) .  \label{4.018}
\end{equation}%
Hence, we define the finite difference derivative $\partial _{x_{i}}$ of the
function $P^{h}$ at the point (\ref{4.018}) 
\begin{equation}
\left. 
\begin{array}{c}
\partial _{x_{i}}P^{h}\left( x_{1,j_{1}},...,x_{i,j_{i}},...,x_{\left(
n-1\right) ,j_{n-1}},z,\alpha \right) =\left( 2h\right) ^{-1}\times \\ 
\\ 
\times \left( 
\begin{array}{c}
P^{h}\left( x_{1,j_{1}},...,x_{i,j_{i}+1},...,x_{\left( n-1\right)
,j_{n-1},}z,\alpha \right) - \\ 
-P^{h}\left( x_{1,j_{1}},...,x_{i,j_{i}-1},...,x_{\left( n-1\right)
,j_{n-1},}z,\alpha \right)%
\end{array}%
\right) , \\ 
\\ 
\forall j_{1},...,j_{i},...j_{n-1}\in \left[ 1,k-1\right] .%
\end{array}%
\right.  \label{4.18}
\end{equation}%
In fact, (\ref{4.18}) is an analog of the following well known formula for
the finite difference derivative of an appropriate function $f\left(
x\right) ,$ $x\in \mathbb{R}:$ 
\begin{equation*}
f^{^{\prime }}\left( x\right) \approx \frac{f\left( x+h\right) -f\left(
x-h\right) }{2h}.
\end{equation*}%
In (\ref{4.18}) $x_{i,j_{i}}+h=x_{i.j_{i}+1}$ and $%
x_{i,j_{i}}-h=x_{i.j_{i}-1}.$

Naturally, the continuous $z,\alpha -$derivatives of the function $P^{h}(%
\overline{x}^{h},z,\alpha )$ is defined in the conventional manner. Using (%
\ref{4.17}), denote 
\begin{equation}
u^{h}\left( x^{h},\alpha \right) =u\left( \overline{x}^{h},z,\alpha \right)
=u^{h}\left( x^{h},\alpha \right) .  \label{4.19}
\end{equation}%
Hence, (\ref{4.18}) and (\ref{4.19}) imply that equation (\ref{4.100}) in
partial finite differences becomes%
\begin{equation}
\left. 
\begin{array}{c}
\partial _{z}u_{\alpha }^{h}+\sum\limits_{i=1}^{n-1}q_{i}^{h}\partial
_{x_{i}}u_{\alpha }^{h}+q_{0}^{h}u_{\alpha }+ \\ 
\\ 
+\sum\limits_{i=1}^{n-1}\partial _{\alpha }q_{i}^{h}\partial
_{x_{i}}u^{h}+\partial _{\alpha }q_{0}u^{h}=0,\text{ } \\ 
\forall \left( x_{1,j_{1}}...,x_{i,j_{i}},...,x_{\left( n-1\right)
,j_{n-1}},z,\alpha \right) =\left( x^{h},\alpha \right) \in \Omega
^{h}\times \left( -A,A\right) .%
\end{array}%
\right.  \label{4.20}
\end{equation}%
Using (\ref{4.10}), (\ref{4.010}), (\ref{4.16}) and (\ref{4.19}), we obtain
the following boundary conditions for the function $u^{h}\left( x^{h},\alpha
\right) :$%
\begin{equation}
\left. 
\begin{array}{c}
u^{h}\left( \overline{x}^{h},B,\alpha \right) =g^{h}\left( \overline{x}%
^{h},\alpha \right) ,\text{ }\left( \overline{x}^{h}\mathbf{,}B,\alpha
\right) \in \partial _{1}\Omega ^{h}\times \left( -A,A\right) , \\ 
\\ 
u^{h}\left( \overline{x}^{h},0,\alpha \right) =0,\text{ }\left( \overline{x}%
^{h}\mathbf{,}0,\alpha \right) \in \partial _{2}\Omega ^{h}\times \left(
-A,A\right) , \\ 
\\ 
u^{h}\left( x^{h},\alpha \right) =0,\text{ }\forall \left( x^{h},\alpha
\right) \in \partial _{3}\Omega ^{h}\times \left( -A,A\right) .%
\end{array}%
\right. \text{ }  \label{4.21}
\end{equation}

Thus, we have obtained equation (\ref{4.20}) in finite differences with
boundary conditions (\ref{4.21}). It follows from (\ref{4.8}) and (\ref{4.19}%
) that the semi-discrete analog $\widetilde{m}^{h}\left( x^{h}\right) $ of
the function $\widetilde{m}\left( x\right) $ is:%
\begin{equation}
\widetilde{m}^{h}\left( x^{h}\right) =\frac{1}{2A}\int\limits_{-A}^{A}%
\left( \partial _{z}u^{h}\left( x^{h},\alpha \right)
+\sum\limits_{i=1}^{n-1}q_{i}^{h}\left( x^{h},\alpha \right)
u_{x_{i}}^{h}+q_{0}^{h}\left( x^{h},\alpha \right) u^{h}\left( x^{h},\alpha
\right) \right) d\alpha .  \label{4.22}
\end{equation}%
We have taken in (\ref{4.22}) the average over $\alpha \in \left(
-A,A\right) $ for the sake of definiteness since the right hand side of
equation (\ref{4.8}) is independent on $\alpha ,$ whereas its left hand side
depends on $\alpha .$

{Body Math}
We now introduce some semi-discrete analogs of $L_{2}-$spaces. By (\ref{2.4}%
) and (\ref{4.13})-(\ref{4.15}) it is natural to introduce the following
semi-discrete analogs $L_{2}^{h}\left( \Omega ^{h}\right) $ and $%
H_{z}^{1,h}\left( \Omega ^{h}\right) $ of the spaces $L_{2}\left( \Omega
\right) $ and $H^{1}\left( \Omega \right) $ respectively:%
\begin{equation}
\left. L_{2}^{h}\left( \Omega ^{h}\right) =\left\{ 
\begin{array}{c}
Y^{h}\left( \overline{x}^{h},z\right) : \\ 
\\ 
\left\Vert Y^{h}\left( \overline{x}^{h},z\right) \right\Vert
_{L_{2}^{h}\left( \Omega ^{h}\right) }=\left( h^{n-1}\sum\limits_{\left(
i,j\right) =\left( 1,1\right) }^{\left( n-1,k-1\right)
}\int\limits_{0}^{B}\left( Y^{h}\left( x_{ij},z\right) \right)
^{2}dz\right) ^{1/2}<\infty 
\end{array}%
\right\} ,\right.   \label{4.23}
\end{equation}%
\begin{equation}
\left. H_{z}^{1,h}\left( \Omega ^{h}\right) =\left\{ 
\begin{array}{c}
Y^{h}\left( \overline{x}^{h},z\right) : \\ 
\\ 
\left\Vert Y^{h}\left( \overline{x}^{h},z\right) \right\Vert
_{H_{z}^{1,h}\left( \Omega ^{h}\right) }= \\ 
\\ 
=\left( h^{n-1}\sum\limits_{\left( i,j\right) =\left( 1,1\right) }^{\left(
n-1,k-1\right) }\int\limits_{0}^{B}\left( \left( \partial _{z}Y^{h}\left(
x_{ij},z\right) \right) ^{2}+\left( Y^{h}\left( x_{ij},z\right) \right)
^{2}\right) dz\right) ^{1/2}<\infty 
\end{array}%
\right\} .\right.   \label{4.230}
\end{equation}
Let $N>1$ be an integer. Based on (\ref{4.23}) and (\ref{4.230}), we
introduce two similar spaces of semi-discrete $N-$D vector functions,%
\begin{equation}
\left. L_{2,N}^{h}\left( \Omega ^{h}\right) =\left\{ 
\begin{array}{c}
R^{h}\left( x^{h}\right) =\left( R_{0}^{h},...,R_{N-1}^{h}\right) ^{T}\left(
x^{h}\right) : \\ 
\\ 
\left\Vert R^{h}\left( x^{h}\right) \right\Vert _{L_{2,N}^{h}\left( \Omega
^{h}\right) }=\left( \sum\limits_{j=0}^{N-1}\left\Vert R_{j}\left(
x^{h}\right) \right\Vert _{L_{2}^{h}\left( \Omega ^{h}\right) }^{2}\right)
^{1/2}<\infty 
\end{array}%
\right\} ,\right.   \label{4.24}
\end{equation}%
\begin{equation}
\left. H_{z,N}^{1,h}\left( \Omega ^{h}\right) =\left\{ 
\begin{array}{c}
R^{h}\left( x^{h}\right) =\left( R_{0}^{h},...,R_{N-1}^{h}\right) ^{T}\left(
x^{h}\right) : \\ 
\\ 
\left\Vert R^{h}\left( x^{h}\right) \right\Vert _{H_{z,N}^{1,h}\left( \Omega
^{h}\right) }=\left( \sum\limits_{j=0}^{N-1}\left\Vert R_{j}\left(
x^{h}\right) \right\Vert _{H_{z,N}^{1,h}\left( \Omega ^{h}\right)
}^{2}\right) ^{1/2}<\infty 
\end{array}%
\right\} .\right.   \label{4.240}
\end{equation}%
Next, using (\ref{2.28}) and (\ref{4.16}), we introduce the discrete analog $%
L_{2,N}^{h}\left( \partial _{1}\Omega ^{h}\right) $ of the space $%
L_{2}\left( \partial _{1}\Omega \right) $%
\begin{equation}
\left. L_{2,N}^{h}\left( \partial _{1}\Omega ^{h}\right) =\left\{ 
\begin{array}{c}
Z^{h}\left( \overline{x}^{h}\right) =\left( z_{0}^{h},...,z_{N-1}^{h}\right)
^{T}\left( \overline{x}^{h}\right) : \\ 
\\ 
\left\Vert Z^{h}\left( \overline{x}^{h}\right) \right\Vert
_{L_{2,N}^{h}\left( \partial _{1}\Omega ^{h}\right) }=\left(
h^{n-1}\sum\limits_{r=0}^{N-1}\sum\limits_{\left( i,j\right) =\left(
1,1\right) }^{\left( n-1,k-1\right) }\left( z^{h}\left( x_{ij}\right)
\right) ^{2}\right) ^{1/2}<\infty 
\end{array}%
\right\} .\right.   \label{4.25}
\end{equation}

Let $\lambda >0$ be a parameter, which we will choose later. We introduce
the Carleman Weight Function $\varphi _{\lambda }\left( z\right) $ as well
as the corresponding semi-discrete Carleman Weighted Space. We will use
these notions in subsection 5.2. We set%
\begin{equation}
\varphi _{\lambda }\left( z\right) =e^{\lambda z}.  \label{4.250}
\end{equation}%
Using (\ref{4.24}), (\ref{4.240}) and (\ref{4.250}), we define two
semi-discrete Carleman Weighted Spaces $L_{2,N,\lambda }^{h}\left( \Omega
^{h}\right) $ and $H_{z,N,\lambda }^{1,h}\left( \Omega ^{h}\right) .$ 
\begin{equation}
\left. L_{2,N,\lambda }^{h}\left( \Omega ^{h}\right) =\left\{ 
\begin{array}{c}
R^{h}\left( x^{h}\right) =\left( R_{0}^{h},...,R_{N-1}^{h}\right) ^{T}\left(
x^{h}\right) : \\ 
\left\Vert R^{h}\right\Vert _{L_{2,N,\lambda }^{h}\left( \Omega ^{h}\right)
}=\left\Vert R^{h}\cdot \varphi _{\lambda }\left( z\right) \right\Vert
_{L_{2,N}^{h}\left( \Omega ^{h}\right) }<\infty 
\end{array}%
\right\} ,\right.   \label{4.251}
\end{equation}%
\begin{equation}
\left. H_{z,N,\lambda }^{1,h}\left( \Omega ^{h}\right) =\left\{ 
\begin{array}{c}
R^{h}\left( x^{h}\right) =\left( R_{0}^{h},...,R_{N-1}^{h}\right) ^{T}\left(
x^{h}\right) : \\ 
\left\Vert R^{h}\right\Vert _{H_{z,N,\lambda }^{1,h}\left( \Omega
^{h}\right) }=\left\Vert R^{h}\cdot \varphi _{\lambda }\left( z\right)
\right\Vert _{H_{z,N}^{1,h}\left( \Omega ^{h}\right) }<\infty 
\end{array}%
\right\} .\right.   \label{4.252}
\end{equation}

\subsection{Representation via the orthonormal basis $\Psi _{j}(\protect%
\alpha )$ in (\protect\ref{3.1})}

\label{sec:4.3}

We assume that the semi-discrete function $u^{h}\left( x^{h},\alpha \right) $
can be represented via the truncated Fourier-like series with respect to the
orthonormal basis (\ref{3.1}),%
\begin{equation}
u^{h}\left( x^{h},\alpha \right) =\sum\limits_{j=0}^{N-1}u_{j}^{h}\left(
x^{h}\right) \Psi _{j}(\alpha ),\text{ }\left( x^{h},\alpha \right) \in 
\overline{\Omega }^{h}\times \left( -A,A\right) .  \label{4.29}
\end{equation}%
Furthermore, we assume that the right hand side of (\ref{4.29}) satisfies
equation (\ref{4.20}), 
\begin{equation}
\left. 
\begin{array}{c}
\sum\limits_{j=0}^{N-1}\partial _{z}u_{j}^{h}\left( x^{h}\right) \Psi
_{j}^{\prime }(\alpha )+\sum\limits_{j=0}^{N-1}\left(
\sum\limits_{i=1}^{n-1}q_{i}^{h}\partial _{x_{i}}u_{j}^{h}\left(
x^{h}\right) \right) \Psi _{j}^{\prime }(\alpha
)+\sum\limits_{j=0}^{N-1}\Psi _{j}^{\prime }(\alpha )q_{0}^{h}u_{j}\left(
x^{h}\right) + \\ 
\\ 
+\sum\limits_{j=0}^{N-1}\Psi _{j}(\alpha )\left(
\sum\limits_{i=1}^{n-1}\partial _{\alpha }q_{i}^{h}\partial
_{x_{i}}u_{j}^{h}\left( x^{h}\right) +\partial _{\alpha }q_{0}\cdot
u_{j}^{h}\left( x^{h}\right) \right) =0,\text{ } \\ 
\\ 
\forall \left( x^{h},\alpha \right) \in \Omega ^{h}\times \left( -A,A\right)
.%
\end{array}%
\right.  \label{4.30}
\end{equation}%
Next, using (\ref{4.21}) and (\ref{4.29}), we obtain the following boundary
conditions for the functions $u_{j}^{h}\left( x^{h}\right) $ 
\begin{equation}
\left. 
\begin{array}{c}
u_{j}^{h}\left( \overline{x}^{h},B\right) =g_{j}^{h}\left( \overline{x}%
^{h}\right) ,\text{ }\left( \overline{x}^{h}\mathbf{,}B\right) \in \partial
_{1}\Omega ^{h},\text{ } \\ 
\\ 
g_{j}^{h}\left( \overline{x}^{h}\right) =\int\limits_{-A}^{A}g^{h}\left( 
\overline{x}^{h},\alpha \right) \Psi _{j}(\alpha )d\alpha ,\text{ } \\ 
\\ 
u_{j}^{h}\left( \overline{x}^{h},0\right) =0,\text{ }\left( \overline{x}^{h}%
\mathbf{,}0\right) \in \partial _{2}\Omega ^{h},\text{ } \\ 
\\ 
u_{j}^{h}\left( x^{h}\right) =0,\text{ }\forall x^{h}\in \partial _{3}\Omega
^{h}, \\ 
\\ 
j=0,...,N-1.%
\end{array}%
\right.  \label{4.31}
\end{equation}%
In addition, we naturally assume that formula (\ref{4.22}) holds for the
function $u^{h}\left( x^{h},\alpha \right) $ being represented via the right
hand side of (\ref{4.29}), i.e.%
\begin{equation}
\widetilde{m}^{h}\left( x^{h}\right) =\frac{1}{2A}\sum\limits_{j=0}^{N-1}%
\int\limits_{-A}^{A}\Psi _{j}(\alpha )\left( \partial _{z}u_{j}^{h}\left(
x^{h}\right) +\sum\limits_{i=1}^{n-1}q_{i}^{h}\partial
_{x_{i}}u_{j}^{h}\left( x^{h}\right) +q_{0}^{h}u_{j}^{h}\left( x^{h}\right)
\right) d\alpha .  \label{4.32}
\end{equation}

Formulas (\ref{4.29})-(\ref{4.32}) complete our transformation procedure.

\textbf{Remark 4.1.} \emph{We do not use below the zero boundary condition
at }$\partial _{2}\Omega ^{h}$\emph{\ in the third line of (\ref{4.31}). The
reason of this is explained in the course of the proof of Theorem 5.2.
Therefore, we ignore below the third line of (\ref{4.31}).}

\section{Lipschitz Stability and Uniqueness}

\label{sec:5}

Introduce the $N-$D vector functions $U^{h}\left( x^{h}\right) ,$ $%
G^{h}\left( \overline{x}^{h}\right) $%
\begin{equation}
U^{h}\left( x^{h}\right) =\left( u_{0}^{h}\left( x^{h}\right)
,...,u_{N-1}^{h}\left( x^{h}\right) \right) ^{T},  \label{5.5}
\end{equation}%
\begin{equation}
G^{h}\left( \overline{x}^{h}\right) =\left( g_{0}^{h}\left( \overline{x}%
^{h}\right) ,...,g_{N-1}^{h}\left( \overline{x}^{h}\right) \right) ^{T}
\label{5.50}
\end{equation}%
where the boundary functions $\left\{ g_{j}^{h}\left( \overline{x}%
^{h}\right) \right\} _{j=0}^{N-1}$ are defined in (\ref{4.31}). The $z-$%
derivative of the vector function $U^{h}\left( x^{h}\right) $ is defined as:%
\begin{equation}
\left. \partial _{z}U^{h}\left( x^{h}\right) =\left( \partial
_{z}u_{0}^{h}\left( x^{h}\right) ,...,\partial _{z}u_{N-1}^{h}\left(
x^{h}\right) \right) ^{T}.\right.  \label{5.6}
\end{equation}

Theorem 5.1 is the main result of this paper.

\textbf{Theorem 5.1.}\emph{\ Assume that conditions (\ref{2.4})-(\ref{2.9}),
(\ref{2.150}), (\ref{4.3}), (\ref{4.5}), (\ref{4.7}), (\ref{4.9})-(\ref%
{4.010}), (\ref{4.13})-(\ref{4.19}), and (\ref{5.5})-\ref{5.6}) hold. Assume
that the semi-discrete function }$u^{h}\left( x^{h},\alpha \right) $\emph{\
has the form (\ref{4.29}), where }$N>1$\emph{\ is an arbitrary fixed integer
and functions }$u_{j}^{h}\in H_{z}^{1,h}\left( \Omega ^{h}\right) ,$ $%
j=0,...,N-1$\emph{. In addition, assume that the semi-discrete function }$%
u^{h}\left( x^{h},\alpha \right) $\emph{\ in the right hand side of (\ref%
{4.29}) satisfies equation (\ref{4.30}) as well as boundary conditions (\ref%
{4.31}) at }$\partial _{1}\Omega ^{h}$ \emph{and} $\partial _{3}\Omega ^{h}$ 
\emph{(see Remark 4.1). Suppose that the semi-discrete analog }$\widetilde{m}%
^{h}\left( x^{h}\right) $\emph{\ of the function }$\widetilde{m}\left(
x\right) $\emph{\ in (\ref{4.1}) has the form (\ref{4.32}). Then there
exists a number }$C=C\left( n,\overline{m},b,\Omega ,h_{0},k,N\right) >0$%
\emph{\ depending only on listed parameters such that the following
Lipschitz stability estimates are valid for the vector function }$%
U^{h}\left( x^{h}\right) $\emph{\ and the function }$\widetilde{m}^{h}\left(
x^{h}\right) :$%
\begin{equation}
\left\Vert U^{h}\right\Vert _{H_{z,N}^{1,h}\left( \Omega ^{h}\right) }\leq
C\left\Vert G^{h}\right\Vert _{L_{2,N}^{h}\left( \partial _{1}\Omega
^{h}\right) },  \label{5.1}
\end{equation}%
\begin{equation}
\left\Vert \widetilde{m}^{h}\right\Vert _{L_{2}^{h}\left( \Omega ^{h}\right)
}\leq C\left\Vert G^{h}\right\Vert _{L_{2,N}^{h}\left( \partial _{1}\Omega
^{h}\right) },  \label{5.2}
\end{equation}%
\emph{where the spaces }$L_{2,N}^{h}\left( \Omega ^{h}\right) $\emph{, }$%
H_{z,N}^{1,h}\left( \Omega ^{h}\right) $ \emph{and} $L_{2,N}^{h}\left(
\partial _{1}\Omega ^{h}\right) $\emph{\ are the ones introduced in (\ref%
{4.24}), (\ref{4.240}) and (\ref{4.25}) respectively. In particular, if all
functions }$g_{j}^{h}\left( \overline{x}^{h}\right) =0$\emph{\ in }$\partial
_{1}\Omega ^{h},$\emph{\ then all (\ref{5.1}) and (\ref{5.2}) imply that all
functions }$u_{j}^{h}\left( x^{h},\alpha \right) =0$\emph{\ in }$\Omega ^{h}$%
\emph{\ and also }$\widetilde{m}^{h}\left( x^{h}\right) =0$\emph{\ in }$%
\Omega ^{h}.$\emph{\ In other words, problem (\ref{4.29})-(\ref{4.32}) has
at most one solution with the properties listed above in this theorem.}

We assume everywhere below without further mentioning that conditions
formulated in the first sentence of Theorem 5.1 hold. In particular, this
means the existence of a number $D=D\left( n,\overline{m},b,\Omega \right)
>0 $ such that (\ref{4.11}) holds.

To prove Theorem 5.1, we first transform in subsection 5.1 equation (\ref%
{4.30}) into a system of $N$ coupled equations with respect to the functions 
$u_{j}\left( x^{h}\right) .$ Next, we prove a new Carleman estimate in
subsection 5.2. Finally, we finish the proof in subsection 5.3. Below $%
C=C\left( n,\overline{m},b,\Omega ,h_{0},k,N\right) >0$ denotes different
numbers depending on these parameters.

\subsection{Beginning of the proof of Theorem 5.1}

\label{sec:5.1}

Let $r\in \left[ 0,N-1\right] $ be an arbitrary integer. Multiply both sides
of equation (\ref{4.30}) by the function $\Psi _{r}\left( \alpha \right) $
and integrate it then with respect to $\alpha \in \left( -A,A\right) .$ We
obtain%
\begin{equation}
\left. 
\begin{array}{c}
\sum\limits_{j=0}^{N-1}\partial _{z}u_{j}^{h}\left( x^{h}\right)
\int\limits_{-A}^{A}\Psi _{j}^{\prime }(\alpha )\Psi _{r}(\alpha )d\alpha +
\\ 
\\ 
+\int\limits_{-A}^{A}\left[ \sum\limits_{j=0}^{N-1}\left(
\sum\limits_{i=1}^{n-1}q_{i}^{h}\cdot \partial _{x_{i}}u_{j}^{h}\left(
x^{h}\right) \right) \Psi _{j}^{\prime }(\alpha
)+\sum\limits_{j=0}^{N-1}\Psi _{j}^{\prime }(\alpha )q_{0}^{h}\cdot
u_{j}^{h}\left( x^{h}\right) \right] \Psi _{r}(\alpha )d\alpha + \\ 
\\ 
+\int\limits_{-A}^{A}\left[ \sum\limits_{j=0}^{N-1}\Psi _{j}(\alpha
)\left( \sum\limits_{i=1}^{n-1}\partial _{\alpha }q_{i}^{h}\cdot \partial
_{x_{i}}u_{j}^{h}\left( x^{h}\right) +\partial _{\alpha }q_{0}\cdot
u_{j}^{h}\left( x^{h}\right) \right) \right] \Psi _{r}(\alpha )d\alpha =0,%
\text{ } \\ 
\\ 
\forall r=0,...,N-1,\text{ }\forall x^{h}\in \Omega ^{h}.%
\end{array}%
\right.  \label{5.3}
\end{equation}%
By (\ref{3.2}) 
\begin{equation}
\int\limits_{-A}^{A}\Psi _{j}^{\prime }(\alpha )\Psi _{r}(\alpha )d\alpha
=a_{jr},  \label{5.4}
\end{equation}%
where the numbers $\left\{ a_{jr}\right\} $ form the matrix $M_{N}$ in (\ref%
{3.4}). Hence, (\ref{3.4}), (\ref{5.5})- (\ref{5.6}), (\ref{5.4}) and Remark
4.1 imply that the system of equations (\ref{5.3}) supplemented by boundary
conditions (\ref{4.31}) can be rewritten as:%
\begin{equation}
\left. 
\begin{array}{c}
M_{N}\cdot \partial _{z}U^{h}\left( x^{h}\right) +F\left(
U_{x_{1}}^{h}\left( x^{h}\right) ,...,U_{x_{n-1}}^{h}\left( x^{h}\right)
,U^{h}\left( x^{h}\right) ,x^{h}\right) =0,\text{ } \\ 
\\ 
U^{h}\left( \overline{x}^{h},B,\alpha \right) =G^{h}\left( \overline{x}%
^{h}\right) ,\text{ }\left( \overline{x}^{h}\mathbf{,}B,\alpha \right) \in
\partial _{1}\Omega ^{h}, \\ 
\\ 
U^{h}\left( x^{h}\right) =0,\text{ }x^{h}\in \partial _{3}\Omega ^{h},%
\end{array}%
\right.  \label{5.7}
\end{equation}%
where the components of the $n-$D vector function $F=\left(
F_{0},...,F_{N-1}\right) ^{T}$ are:%
\begin{equation}
\left. 
\begin{array}{c}
F_{r}\left( U_{x_{1}}^{h}\left( x^{h}\right) ,...,U_{x_{n-1}}^{h}\left(
x^{h}\right) ,U^{h}\left( x^{h}\right) ,x^{h}\right) = \\ 
\\ 
=\int\limits_{-A}^{A}\left[ \sum\limits_{j=0}^{N-1}\left(
\sum\limits_{i=1}^{n-1}q_{i}^{h}\partial _{x_{i}}u_{j}^{h}\left(
x^{h}\right) \right) \Psi _{j}^{\prime }(\alpha
)+\sum\limits_{j=0}^{N-1}\Psi _{j}^{\prime }(\alpha
)q_{0}^{h}u_{j}^{h}\left( x^{h}\right) \right] \Psi _{r}(\alpha )d\alpha +
\\ 
\\ 
+\int\limits_{-A}^{A}\left[ \sum\limits_{j=0}^{N-1}\Psi _{j}(\alpha
)\left( \sum\limits_{i=1}^{n-1}\partial _{\alpha }q_{i}^{h}\partial
_{x_{i}}u_{j}^{h}\left( x^{h}\right) +\partial _{\alpha
}q_{0}u_{j}^{h}\left( x^{h}\right) \right) \right] \Psi _{r}(\alpha )d\alpha
, \\ 
\\ 
r=0,...,N-1.%
\end{array}%
\right.  \label{5.8}
\end{equation}%
Since the matrix $M_{N}$ is invertible by Theorem 3.1, then multiplying from
the left both parts of the matrix equation in the first line of (\ref{5.7})
by the matrix $M_{N}^{-1},$ we obtain the following system of PDEs in
partial finite differences%
\begin{equation}
\partial _{z}U^{h}\left( x^{h}\right) +M_{N}^{-1}F\left( U_{x_{1}}^{h}\left(
x^{h}\right) ,...,U_{x_{n-1}}^{h}\left( x^{h}\right) ,U^{h}\left(
x^{h}\right) ,x^{h}\right) =0,\text{ }x^{h}\in \Omega ^{h}.  \label{5.9}
\end{equation}%
We now turn our attention to a new Carleman estimate.

\subsection{A new Carleman estimate}

\label{sec:5.2}

Let 
\begin{equation}
V^{h}\left( x^{h}\right) =\left( v_{0}^{h}\left( x^{h}\right)
,...,v_{N-1}^{h}\left( x^{h}\right) \right) ^{T}  \label{5.10}
\end{equation}%
be an arbitrary semi-discrete vector function is such that 
\begin{equation}
V^{h},\partial _{z}V^{h}\in L_{2}^{h}\left( \Omega ^{h}\right) .
\label{5.11}
\end{equation}%
Recalling the third line of (\ref{5.7}), assume that 
\begin{equation}
V^{h}\left( x^{h}\right) =0,\text{ }x^{h}\in \partial _{3}\Omega ^{h}.
\label{5.12}
\end{equation}

There is a peculiarity in the proof of Theorem 5.2. This peculiarity is
linked with boundary condition (\ref{5.12}).

\textbf{Theorem 5.2} (Carleman estimate). \emph{Let }%
\begin{equation*}
F=\left( F_{0},...,F_{N-1}\right) ^{T}\left( V_{x_{1}}^{h}\left(
x^{h}\right) ,...,V_{x_{n-1}}^{h}\left( x^{h}\right) ,V^{h}\left(
x^{h}\right) ,x^{h}\right)
\end{equation*}%
\emph{be the }$N-$\emph{D vector function with its components defined in (%
\ref{5.8}) in which components of the vector function }$U^{h}\left(
x^{h}\right) $\emph{\ in (\ref{5.5}) are replaced with components of the
vector function }$V^{h}\left( x^{h}\right) $ \emph{defined in} \emph{(\ref%
{5.10}). Let\ }$\varphi _{\lambda }\left( z\right) $ \emph{be the Carleman
Weight Function defined in (\ref{4.250}), and let }$L_{2,N,\lambda
}^{h}\left( \Omega ^{h}\right) $ \emph{and} $H_{z,N,\lambda }^{1,h}\left(
\Omega ^{h}\right) $\emph{\ be the semi-discrete Carleman Weighted Spaces
defined in (\ref{4.251}) and (\ref{4.252}) respectively.\ Then there exists
a sufficiently large number }$\lambda _{0}=\lambda _{0}\left( n,\overline{m}%
,b,\Omega ,h_{0},k,N\right) >1$\emph{\ depending only on listed parameters
such that the following Carleman estimate is valid for all vector functions }%
$V^{h}\left( x^{h}\right) $\emph{\ in (\ref{5.10}) with the property (\ref%
{5.11}) satisfying boundary conditions (\ref{5.12})}%
\begin{equation}
\left. 
\begin{array}{c}
\left\Vert \partial _{z}V^{h}\left( x^{h}\right) +M_{N}^{-1}F\left(
V_{x_{1}}^{h}\left( x^{h}\right) ,...,V_{x_{n-1}}^{h}\left( x^{h}\right)
,V^{h}\left( x^{h}\right) ,x^{h}\right) \right\Vert _{L_{2,N,\lambda
}^{h}\left( \Omega ^{h}\right) }^{2}\geq \\ 
\\ 
\geq \left( 1/8\right) \left\Vert V^{h}\left( x^{h}\right) \right\Vert
_{H_{z,N,\lambda }^{1,h}\left( \Omega ^{h}\right) }^{2}-\lambda e^{2\lambda
B}\left\Vert V^{h}\left( x^{h}\right) \right\Vert _{L_{2,N}^{h}\left(
\partial _{1}\Omega ^{h}\right) }^{2}, \\ 
\\ 
\forall \lambda \geq \lambda _{0}.%
\end{array}%
\right.  \label{5.16}
\end{equation}

\textbf{Proof}. By Cauchy-Schwarz inequality%
\begin{equation}
\left. 
\begin{array}{c}
\left\Vert \partial _{z}V^{h}\left( x^{h}\right) +M_{N}^{-1}F\left(
V_{x_{1}}^{h}\left( x^{h}\right) ,...,V_{x_{n-1}}^{h}\left( x^{h}\right)
,V^{h}\left( x^{h}\right) ,x^{h}\right) \right\Vert _{L_{2,N,\lambda
}^{h}\left( \Omega ^{h}\right) }^{2}\geq  \\ 
\\ 
\geq \left( 1/2\right) \left\Vert \partial _{z}V^{h}\left( x^{h}\right)
\right\Vert _{L_{2,N,\lambda }^{h}\left( \Omega ^{h}\right) }^{2}- \\ 
\\ 
-C\left\Vert M_{N}^{-1}F\left( V_{x_{1}}^{h}\left( x^{h}\right)
,...,V_{x_{n-1}}^{h}\left( x^{h}\right) ,V^{h}\left( x^{h}\right)
,x^{h}\right) \right\Vert _{L_{2,N,\lambda }^{h}\left( \Omega ^{h}\right)
}^{2}.%
\end{array}%
\right.   \label{5.17}
\end{equation}%
It follows from (\ref{4.11}), (\ref{5.5}), (\ref{5.8}), (\ref{5.10}) and (%
\ref{5.11}) that%
\begin{equation*}
\left. 
\begin{array}{c}
\left\Vert M_{N}^{-1}F\left( V_{x_{1}}^{h}\left( x^{h}\right)
,...,V_{x_{n-1}}^{h}\left( x^{h}\right) ,V^{h}\left( x^{h}\right)
,x^{h}\right) \right\Vert _{L_{2,N,\lambda }^{h}\left( \Omega ^{h}\right)
}^{2}\leq  \\ 
\\ 
\leq C\sum\limits_{i=1}^{n-1}\left\Vert V_{x_{i}}^{h}\left( x^{h}\right)
\right\Vert _{L_{2,N,\lambda }^{h}\left( \Omega ^{h}\right)
}^{2}+C\left\Vert V^{h}\left( x^{h}\right) \right\Vert _{L_{2,N,\lambda
}^{h}\left( \Omega ^{h}\right) }^{2}.%
\end{array}%
\right. 
\end{equation*}%
Hence, using (\ref{5.17}), we obtain%
\begin{equation}
\left. 
\begin{array}{c}
\left\Vert \partial _{z}V^{h}\left( x^{h}\right) +M_{N}^{-1}F\left(
V_{x_{1}}^{h}\left( x^{h}\right) ,...,V_{x_{n-1}}^{h}\left( x^{h}\right)
,V^{h}\left( x^{h}\right) ,x^{h}\right) \right\Vert _{L_{2,N,\lambda
}^{h}\left( \Omega ^{h}\right) }^{2}\geq  \\ 
\\ 
\geq \left( 1/2\right) \left\Vert \partial _{z}V^{h}\left( x^{h}\right)
\right\Vert _{L_{2,N,\lambda }^{h}\left( \Omega ^{h}\right) }^{2}- \\ 
\\ 
-C\sum\limits_{i=1}^{n-1}\left\Vert V_{x_{i}}^{h}\left( x^{h}\right)
\right\Vert _{L_{2,N,\lambda }^{h}\left( \Omega ^{h}\right)
}^{2}-C\left\Vert V^{h}\left( x^{h}\right) \right\Vert _{L_{2,N,\lambda
}^{h}\left( \Omega ^{h}\right) }^{2}.%
\end{array}%
\right.   \label{5.18}
\end{equation}

We now consider the term $\left\Vert V_{x_{i}}^{h}\left( x^{h}\right)
\right\Vert _{L_{2,,N}^{h}\left( \Omega ^{h}\right) }^{2}$ and address the
above mentioned peculiarity linked with the second line of (\ref{5.12}).
Consider, for example, the function $\partial _{x_{1}}v_{0}^{h}\left(
x^{h}\right) .$ It follows from (\ref{4.18}) that for any point $\left(
x_{1,j_{1}},...,x_{\left( n-1\right) ,j_{n-1}},z\right) \in \Omega ^{h}$ 
\begin{equation*}
\left. 
\begin{array}{c}
\partial _{x_{1}}v_{0}^{h}\left( x_{1,j_{1}},...,x_{\left( n-1\right)
,j_{n-1}},z\right) =\left( 2h\right) ^{-1}\times  \\ 
\\ 
\times \left( v_{0}^{h}\left( x_{1,j_{1}+1},...,x_{\left( n-1\right)
,j_{n-1}},z\right) -v_{0}^{h}\left( x_{1,j_{1}-1},...,x_{\left( n-1\right)
,j_{n-1}}z\right) \right) , \\ 
\\ 
\forall j_{1},...,j_{n-1}\in \left[ 1,k-1\right] .%
\end{array}%
\right. 
\end{equation*}%
Hence, using (\ref{4.14}), we obtain%
\begin{equation}
\left. 
\begin{array}{c}
\left[ \partial _{x_{1}}v_{0}^{h}\left( x_{1,j_{1}},...,x_{\left( n-1\right)
,j_{n-1}},z\right) \right] ^{2}\leq  \\ 
\\ 
\leq \left( 2h_{0}^{2}\right) ^{-1}\left\{ \left[ v_{0}^{h}\left(
x_{1,j_{1}+1},...,x_{\left( n-1\right) ,j_{n-1}},z\right) \right] ^{2}+\left[
v_{0}^{h}\left( x_{1,j_{1}-1},...,x_{\left( n-1\right) ,j_{n-1}}z\right) %
\right] ^{2}\right\} , \\ 
\\ 
\forall j_{1},...,j_{n-1}\in \left[ 1,k-1\right] .%
\end{array}%
\right.   \label{5.19}
\end{equation}%
Consider now the following sum:%
\begin{equation}
\sum\limits_{\left( j_{2},...,j_{n-1}\right) =\left( 1,...,1\right)
}^{\left( k-1,...,k-1\right)
}\sum\limits_{j_{1}=1}^{k-1}\int\limits_{0}^{B}\left[ v_{0}^{h}\left(
x_{1,j_{1}+1},...,x_{\left( n-1\right) ,j_{n-1}},z\right) \right]
^{2}\varphi _{\lambda }^{2}\left( z\right) dz  \label{5.20}
\end{equation}%
It follows from (\ref{4.13}), (\ref{4.15}) and (\ref{5.12}) that 
\begin{equation*}
v_{0}^{h}\left( x_{1,k},...,x_{\left( n-1\right) ,j_{n-1}},z\right) =0.
\end{equation*}%
Hence, denoting in (\ref{5.20}) $j_{1}^{\prime }=j_{1}+1$, keeping the same
notation for brevity and using (\ref{5.14}), we obtain%
\begin{equation}
\left. 
\begin{array}{c}
\sum\limits_{\left( j_{2},...,j_{n-1}\right) =\left( 1,...,1\right)
}^{\left( k-1,...,k-1\right)
}\sum\limits_{j_{1}=1}^{k-1}\int\limits_{0}^{B}\left[ v_{0}^{h}\left(
x_{1,j_{1}+1},...,x_{\left( n-1\right) ,j_{n-1}},z\right) \right]
^{2}\varphi _{\lambda }^{2}\left( z\right) dz= \\ 
\\ 
=\sum\limits_{\left( j_{2},...,j_{n-1}\right) =\left( 1,...,1\right)
}^{\left( k-1,...,k-1\right) }\sum\limits_{j_{1}^{\prime
}=2}^{k-1}\int\limits_{0}^{B}\left[ v_{0}^{h}\left( x_{1,j_{1}^{\prime
}},...,x_{\left( n-1\right) ,j_{n-1}},z\right) \right] ^{2}\varphi _{\lambda
}^{2}\left( z\right) dz\leq  \\ 
\\ 
\leq \sum\limits_{\left( j_{1},,...,j_{n-1}\right) =\left( 1,...,1\right)
}^{\left( k-1,...,k-1\right) }\int\limits_{0}^{B}\left[ v_{0}^{h}\left(
x_{1,j_{1}},...,x_{\left( n-1\right) ,j_{n-1}},z\right) \right] ^{2}\varphi
_{\lambda }^{2}\left( z\right) dz= \\ 
\\ 
=\left\Vert v_{0}^{h}\left( x^{h}\right) \right\Vert _{L_{2,\lambda
}^{h}\left( \Omega ^{h}\right) }^{2}.%
\end{array}%
\right.   \label{5.21}
\end{equation}%
Similarly we obtain for the second term in the second line of (\ref{5.19})%
\begin{equation}
\left. 
\begin{array}{c}
\sum\limits_{\left( j_{2},...,j_{n-1}\right) =\left( 1,...,1\right)
}^{\left( k-1,...,k-1\right)
}\sum\limits_{j_{1}=1}^{k-1}\int\limits_{0}^{B}\left[ v_{0}^{h}\left(
x_{1,j_{1}+1},...,x_{\left( n-1\right) ,j_{n-1}},z\right) \right]
^{2}\varphi _{\lambda }^{2}\left( z\right) dz\leq  \\ 
\\ 
\leq \left\Vert v_{0}^{h}\left( x^{h}\right) \right\Vert _{L_{2,\lambda
}^{h}\left( \Omega ^{h}\right) }^{2}.%
\end{array}%
\right.   \label{5.22}
\end{equation}%
Thus, (\ref{5.19}), (\ref{5.21}) and (\ref{5.22}) imply%
\begin{equation}
\left\Vert \partial _{x_{1}}v_{0}^{h}\left( x^{h}\right) \right\Vert
_{L_{2,\lambda }^{h}\left( \Omega ^{h}\right) }^{2}\leq C\left\Vert
v_{0}^{h}\left( x^{h}\right) \right\Vert _{L_{2,\lambda }^{h}\left( \Omega
^{h}\right) }^{2}.  \label{5.23}
\end{equation}%
Estimate (\ref{5.22}) addresses the above mentioned peculiarity. Obviously,
estimates for Carleman weighted norms 
\begin{equation*}
\left\Vert \partial _{x_{i}}v_{j}^{h}\left( x^{h}\right) \right\Vert
_{L_{2,\lambda }^{h}\left( \Omega ^{h}\right) }^{2},i=1,...,n-1,j=0,...,N-1
\end{equation*}%
are completely similar with (\ref{5.23}). Hence,%
\begin{equation}
\left\Vert V_{x_{i}}^{h}\left( x^{h}\right) \right\Vert _{L_{2,N,\lambda
}^{h}\left( \Omega ^{h}\right) }^{2}\leq C\left\Vert V^{h}\left(
x^{h}\right) \right\Vert _{L_{2,N,\lambda }^{h}\left( \Omega ^{h}\right)
}^{2},\text{ }i=1,...,n-1.  \label{5.230}
\end{equation}%
Hence, recalling that $C=C\left( n,\overline{m},b,\Omega ,h_{0},k,N\right) >0
$ denotes different numbers depending only on listed parameters, we obtain
that it follows from (\ref{5.230}) that the third line of (\ref{5.18}) can
be estimated as:%
\begin{equation}
\left. 
\begin{array}{c}
-C\sum\limits_{i=1}^{n-1}\left\Vert V_{x_{i}}^{h}\left( x^{h}\right)
\right\Vert _{L_{2,N,\lambda }^{h}\left( \Omega ^{h}\right)
}^{2}-C\left\Vert V^{h}\left( x^{h}\right) \right\Vert _{L_{2,N,\lambda
}^{h}\left( \Omega ^{h}\right) }^{2}\geq  \\ 
\geq -C_{1}\left\Vert V^{h}\left( x^{h}\right) \right\Vert _{L_{2,N,\lambda
}^{h}\left( \Omega ^{h}\right) }^{2},%
\end{array}%
\right.   \label{5.24}
\end{equation}%
where we temporary use $C_{1}=C_{1}\left( n,\overline{m},b,\Omega
,h_{0},k,N\right) >0:$ to avoid a confusion. Combining (\ref{5.17}) with (%
\ref{5.24}) and using $C$ again, we obtain%
\begin{equation}
\left. 
\begin{array}{c}
\left\Vert \partial _{z}V^{h}\left( x^{h}\right) +M_{N}^{-1}F\left(
V_{x_{1}}^{h}\left( x^{h}\right) ,...,V_{x_{n-1}}^{h}\left( x^{h}\right)
,V^{h}\left( x^{h}\right) ,x^{h}\right) \right\Vert _{L_{2,N,\lambda
}^{h}\left( \Omega ^{h}\right) }^{2}\geq  \\ 
\\ 
\geq \left( 1/2\right) \left\Vert \partial _{z}V^{h}\left( x^{h}\right)
\right\Vert _{L_{2,N,\lambda }^{h}\left( \Omega ^{h}\right)
}^{2}-C\left\Vert V^{h}\left( x^{h}\right) \right\Vert _{L_{2,N,\lambda
}^{h}\left( \Omega ^{h}\right) }^{2}.%
\end{array}%
\right.   \label{5.25}
\end{equation}

We have not yet used any properties of the Carleman Weight Function $\varphi
_{\lambda }\left( z\right) $ in (\ref{4.250}). We are now starting to do
this. More precisely, we now will estimate from the below the term $\left(
1/2\right) \left\Vert \partial _{z}V^{h}\left( x^{h}\right) \right\Vert
_{L_{2,,N,\lambda }^{h}\left( \Omega ^{h}\right) }^{2}$ in (\ref{5.25}).

We use below the scalar product in $\mathbb{R}^{N},$ e.g. $\partial
_{z}V^{h}\left( x^{h}\right) V^{h}\left( x^{h}\right) $ means the scalar
product of vectors $\partial _{z}V^{h}\left( x^{h}\right) ,V^{h}\left(
x^{h}\right) \in \mathbb{R}^{N}$, respectively, e.g. $\left( V^{h}\left(
x^{h}\right) \right) ^{2}=V^{h}\left( x^{h}\right) V^{h}\left( x^{h}\right)
. $

Change variables as%
\begin{equation}
W^{h}\left( x^{h}\right) =V^{h}\left( x^{h}\right) e^{\lambda z}.
\label{5.26}
\end{equation}%
Then%
\begin{equation*}
V^{h}\left( x^{h}\right) =W^{h}\left( x^{h}\right) e^{-\lambda z}.
\end{equation*}%
Hence, 
\begin{equation*}
\partial _{z}V^{h}\left( x^{h}\right) =\left( \partial _{z}W^{h}\left(
x^{h}\right) -\lambda W^{h}\left( x^{h}\right) \right) e^{-\lambda z}.
\end{equation*}%
Hence,%
\begin{equation}
\left( \partial _{z}V^{h}\left( x^{h}\right) \right) ^{2}e^{2\lambda
z}=\left( \partial _{z}W^{h}\left( x^{h}\right) -\lambda W^{h}\left(
x^{h}\right) \right) ^{2}.  \label{5.28}
\end{equation}

We now estimate from the below the right hand side of (\ref{5.28}). We have%
\begin{equation*}
\left. 
\begin{array}{c}
\left( \partial _{z}W^{h}\left( x^{h}\right) -\lambda W^{h}\left(
x^{h}\right) \right) ^{2}= \\ 
\\ 
=\left( \partial _{z}W^{h}\left( x^{h}\right) \right) ^{2}-2\lambda \partial
_{z}W^{h}\left( x^{h}\right) W^{h}\left( x^{h}\right) +\lambda ^{2}\left(
W^{h}\left( x^{h}\right) \right) ^{2}\geq  \\ 
\\ 
=\left( 1/2\right) \left( \partial _{z}W^{h}\left( x^{h}\right) \right)
^{2}+\lambda ^{2}\left( W^{h}\left( x^{h}\right) \right) ^{2}+\partial
_{z}\left( -\lambda \left( W^{h}\left( x^{h}\right) \right) ^{2}\right) .%
\end{array}%
\right. 
\end{equation*}%
Thus, we have established that%
\begin{equation}
\left. 
\begin{array}{c}
\left( \partial _{z}V^{h}\left( x^{h}\right) \right) ^{2}e^{2\lambda z}= \\ 
\\ 
=\left( \partial _{z}W^{h}\left( x^{h}\right) \right) ^{2}+\lambda
^{2}\left( V^{h}\left( x^{h}\right) \right) ^{2}e^{2\lambda z}+\partial
_{z}\left( -\lambda \left( V^{h}\left( x^{h}\right) \right) ^{2}e^{2\lambda
z}\right) .%
\end{array}%
\right.   \label{5.29}
\end{equation}%
Change variables in the first term in the second line of (\ref{5.29}) via
returning again to the function $V^{h}\left( x^{h}\right) $ as in (\ref{5.26}%
). Using Cauchy-Schwarz inequality, we obtain%
\begin{equation*}
\left. 
\begin{array}{c}
\left( \partial _{z}W^{h}\left( x^{h}\right) \right) ^{2}=\left( \partial
_{z}V^{h}\left( x^{h}\right) +\lambda V^{h}\left( x^{h}\right) \right)
^{2}e^{2\lambda z}= \\ 
=\left( \partial _{z}V^{h}\left( x^{h}\right) \right) ^{2}e^{2\lambda
z}+2\lambda \partial _{z}V^{h}\left( x^{h}\right) V^{h}\left( x^{h}\right)
e^{2\lambda z}+\lambda ^{2}\left( V^{h}\left( x^{h}\right) \right)
^{2}e^{2\lambda z}\geq  \\ 
\geq \left( 1/2\right) \left( \partial _{z}V^{h}\left( x^{h}\right) \right)
^{2}e^{2\lambda z}-\lambda ^{2}\left( V^{h}\left( x^{h}\right) \right)
^{2}e^{2\lambda z}.%
\end{array}%
\right. 
\end{equation*}%
Thus, we have established that%
\begin{equation*}
\left( \partial _{z}W^{h}\left( x^{h}\right) \right) ^{2}\geq \frac{1}{2}%
\left( \partial _{z}V^{h}\left( x^{h}\right) \right) ^{2}e^{2\lambda
z}-\lambda ^{2}\left( V^{h}\left( x^{h}\right) \right) ^{2}e^{2\lambda z}.
\end{equation*}%
Dividing this inequality by 2 and combining then the resulting inequality
with (\ref{5.29}), we obtain 
\begin{equation}
\left. 
\begin{array}{c}
\left( \partial _{z}V^{h}\left( x^{h}\right) \right) ^{2}e^{2\lambda z}\geq 
\\ 
\\ 
\geq \left( 1/4\right) \left( \partial _{z}V^{h}\left( x^{h}\right) \right)
^{2}e^{2\lambda z}+\left( 1/2\right) \lambda ^{2}\left( V^{h}\left(
x^{h}\right) \right) ^{2}e^{2\lambda z} \\ 
\\ 
+\partial _{z}\left( -\lambda \left( V^{h}\left( x^{h}\right) \right)
^{2}e^{2\lambda z}\right) .%
\end{array}%
\right.   \label{5.30}
\end{equation}%
Note that 
\begin{equation}
\left. 
\begin{array}{c}
\int\limits_{0}^{B}\partial _{z}\left( -\lambda \left( V^{h}\left(
x^{h}\right) \right) ^{2}e^{2\lambda z}\right) dz=-\lambda \left(
V^{h}\left( \overline{x}^{h},B\right) \right) ^{2}e^{2\lambda B}+\lambda
\left( V^{h}\left( \overline{x}^{h},0\right) \right) ^{2}\geq  \\ 
\\ 
\geq -\lambda \left( V^{h}\left( \overline{x}^{h},B\right) \right)
^{2}e^{2\lambda B}.%
\end{array}%
\right.   \label{5.301}
\end{equation}%
In fact, (\ref{5.301}) is the reason of our Remark 4.1. Thus, combining (\ref%
{5.25}) with (\ref{5.30}) and (\ref{5.301}) and using (\ref{4.25}), we obtain%
\begin{equation}
\left. 
\begin{array}{c}
\left\Vert \partial _{z}V^{h}\left( x^{h}\right) +M_{N}^{-1}F\left(
V_{x_{1}}^{h}\left( x^{h}\right) ,...,V_{x_{n-1}}^{h}\left( x^{h}\right)
,V^{h}\left( x^{h}\right) ,x^{h}\right) \right\Vert _{L_{2,N,\lambda
}^{h}\left( \Omega ^{h}\right) }^{2}\geq  \\ 
\\ 
\geq \left( 1/8\right) \left\Vert \partial _{z}V^{h}\left( x^{h}\right)
\right\Vert _{L_{2,N,\lambda }^{h}\left( \Omega ^{h}\right) }^{2}+\left(
1/2\right) \lambda ^{2}\left\Vert V^{h}\left( x^{h}\right) \right\Vert
_{L_{2,N,\lambda }^{h}\left( \Omega ^{h}\right) }^{2}- \\ 
-C\left\Vert V^{h}\left( x^{h}\right) \right\Vert _{L_{2,N,\lambda
}^{h}\left( \Omega ^{h}\right) }^{2}-\lambda e^{2\lambda B}\left\Vert
V^{h}\left( \overline{x}^{h},B\right) \right\Vert _{L_{2,N}^{h}\left(
\partial _{1}\Omega ^{h}\right) }^{2}.%
\end{array}%
\right.   \label{5.31}
\end{equation}%
Choose a sufficiently large $\lambda _{0}=\lambda _{0}\left( n,\overline{m}%
,b,\Omega ,h_{0},k,N\right) >1$ such that 
\begin{equation*}
\frac{\lambda _{0}^{2}}{4}>C.
\end{equation*}%
Then in (\ref{5.31}) 
\begin{equation}
\left. 
\begin{array}{c}
\left( 1/2\right) \lambda ^{2}\left\Vert V^{h}\left( x^{h}\right)
\right\Vert _{L_{2,N,\lambda }^{h}\left( \Omega ^{h}\right)
}^{2}-C\left\Vert V^{h}\left( x^{h}\right) \right\Vert _{L_{2,N,\lambda
}^{h}\left( \Omega ^{h}\right) }^{2}\geq  \\ 
\\ 
\geq \left( 1/4\right) \lambda ^{2}\left\Vert V^{h}\left( x^{h}\right)
\right\Vert _{L_{2,N,\lambda }^{h}\left( \Omega ^{h}\right) }^{2},\text{ }%
\forall \lambda \geq \lambda _{0}.%
\end{array}%
\right.   \label{5.32}
\end{equation}%
Combining (\ref{5.32}) with (\ref{5.31}), we obtain%
\begin{equation}
\left. 
\begin{array}{c}
\left\Vert \partial _{z}V^{h}\left( x^{h}\right) +M_{N}^{-1}F\left(
V_{x_{1}}^{h}\left( x^{h}\right) ,...,V_{x_{n-1}}^{h}\left( x^{h}\right)
,V^{h}\left( x^{h}\right) ,x^{h}\right) \right\Vert _{L_{2,N,\lambda
}^{h}\left( \Omega ^{h}\right) }^{2}\geq  \\ 
\\ 
\geq \left( 1/8\right) \left\Vert \partial _{z}V^{h}\left( x^{h}\right)
\right\Vert _{L_{2,N,\lambda }^{h}\left( \Omega ^{h}\right) }^{2}+\left(
1/4\right) \lambda ^{2}\left\Vert V^{h}\left( x^{h}\right) \right\Vert
_{L_{2,N,\lambda }^{h}\left( \Omega ^{h}\right) }^{2}- \\ 
\\ 
-\lambda e^{2\lambda B}\left\Vert V^{h}\left( \overline{x}^{h},B\right)
\right\Vert _{L_{2,N}^{h}\left( \partial _{1}\Omega ^{h}\right) }^{2}\geq 
\\ 
\\ 
\geq \left( 1/8\right) \left\Vert V^{h}\left( x^{h}\right) \right\Vert
_{H_{z,N,\lambda }^{1,h}\left( \Omega ^{h}\right) }^{2}-\lambda e^{2\lambda
B}\left\Vert V^{h}\left( \overline{x}^{h},B\right) \right\Vert
_{L_{2,N}^{h}\left( \partial _{1}\Omega ^{h}\right) }^{2}%
\end{array}%
\right.   \label{5.320}
\end{equation}

The target Carleman estimate (\ref{5.16}) of this theorem follows from (\ref%
{5.320}) immediately. \ $\square $

\subsection{Continuation of the proof of Theorem 5.1}

\label{sec:5.3}

By (\ref{5.9}) 
\begin{equation*}
\left\Vert \partial _{z}U^{h}\left( x^{h}\right) +M_{N}^{-1}F\left(
U_{x_{1}}^{h}\left( x^{h}\right) ,...,U_{x_{n-1}}^{h}\left( x^{h}\right)
,U^{h}\left( x^{h}\right) ,x^{h}\right) \right\Vert _{L_{2,N,\lambda
}^{h}\left( \Omega ^{h}\right) }^{2}=0.
\end{equation*}%
Hence, (\ref{4.25}), (\ref{5.5})-(\ref{5.6}) and (\ref{5.16}) imply%
\begin{equation}
\left\Vert U^{h}\left( x^{h}\right) \right\Vert _{H_{z,N,\lambda
}^{1,h}\left( \Omega ^{h}\right) }^{2}\leq 8\lambda e^{2\lambda B}\left\Vert
G^{h}\left( \overline{x}^{h}\right) \right\Vert _{L_{2,N}^{h}\left( \partial
_{1}\Omega ^{h}\right) }^{2}\text{, \ }\forall \lambda \geq \lambda _{0}.
\label{5.33}
\end{equation}%
Since by (\ref{4.250}) $\varphi _{\lambda }\left( z\right) \geq 1$ for $z\in
\left( 0,B\right) ,$ then (\ref{4.240}), (\ref{4.252}) and (\ref{5.33}) imply%
\begin{equation}
\left\Vert U^{h}\left( x^{h}\right) \right\Vert _{H_{z,N}^{1,h}\left( \Omega
^{h}\right) }^{2}\leq 8\lambda e^{2\lambda B}\left\Vert G^{h}\left( 
\overline{x}^{h}\right) \right\Vert _{L_{2,N}^{h}\left( \partial _{1}\Omega
^{h}\right) }^{2}\text{, \ }\forall \lambda \geq \lambda _{0}.  \label{5.34}
\end{equation}%
Choosing in (\ref{5.34}) $\lambda =\lambda _{0},$ we obtain (\ref{5.1}),
which is the first target Lipschitz stability estimate of this theorem. The
second target Lipschitz stability estimate (\ref{5.2}) is obtained from (\ref%
{5.1}) in the obvious manner via using (\ref{4.32}) and (\ref{5.5}). It is
clear from the formulation of Theorem 5.1 that uniqueness follows
immediately from (\ref{5.1}) and (\ref{5.2}). \ \ \ \ $\square $

\textbf{Acknowledgment.} This research was partially supported by the
National Science Foundation grant DMS 2436227.

\textsc{University of North Carolina at Charlotte, Charlotte, NC, 28223, USA}

\emph{E-mail}: mklibanv@charlotte.edu

\end{document}